\documentclass[10pt,conference]{IEEEtran}
\usepackage[noadjust]{cite}      % Written by Donald Arseneau
\usepackage{graphicx}  % Written by David Carlisle and Sebastian Rahtz
\usepackage{amsmath,amssymb,epsfig}
\usepackage{multicol}
\usepackage{algorithm}
\usepackage{url}
\usepackage{float}
\usepackage{amsfonts}
\usepackage{theorem}
\usepackage{algorithm}
\usepackage{url}
\usepackage{float}
\usepackage{amsfonts}
\usepackage{theorem}
\makeatletter
    \newcommand\figcaption{\def\@captype{figure}\caption}
    \newcommand\tabcaption{\def\@captype{table}\caption}
\makeatother

\newtheorem{thm}{Theorem}%[subsection]

\newtheorem{lem}{Lemma}

\theoremstyle{definition}
\newtheorem{defn}{Definition}
\theoremstyle{remark}

\begin{document}

\title{Multilevel Coding for Channels with Non-uniform Inputs
and Rateless Transmission over the BSC}

\author{Jing~Jiang and Krishna~R.~Narayanan\\
Department of Electrical and Computer Engineering, \\
Texas A\&M University,\\
College Station, TX, 77840, U.S.A} \maketitle{}

\begin{abstract}
We consider coding schemes for channels with non-uniform inputs (NUI), where standard linear block codes can not
be applied directly. We show that multilevel coding (MLC) with a set of linear codes and a deterministic mapper
can achieve the information rate of the channel with NUI. The mapper, however, does not have to be one-to-one.
As an application of the proposed MLC scheme, we present a rateless transmission scheme over the binary
symmetric channel (BSC).
\end{abstract}

\section{Introduction}
\label{sec:introduction}

Consider a discrete memoryless channel (DMC) with input $X \in {\cal X}$, output $Y \in {\cal Y}$ and let $P_X$
be the input distribution. When $P_X$ is the uniform distribution on ${\cal X}$, denoted by
$\sf{unif}(\cal{X})$, it is well known that linear codes can be directly used to achieve the information rate
corresponding to $P_X$ \cite{gallager63}. When $|{\cal{X}}| = 2^m$, binary linear codes along with multilevel
coding (MLC) suffice to achieve the information rate corresponding to $\sf{unif}(\cal{X})$. Given extensive
recent results on designing good LDPC codes for binary input symmetric channels \cite{rich01b,richardson_book},
it suffices to say that LDPC codes provide a good practical solution to this communication problem.

However, in many cases, we are interested in input distributions which are not the uniform distribution. This
could be because the capacity achieving distribution is non-uniform (for example, the Z channel
\cite{mceliece_talk}). Or, other signalling constraints may force us to use non-uniform inputs. An example of
this is optical channels with cross-talks, where the probability of 1s transmitted by each user $p_1 = P(X = 1)$
has to be constrained to be $p_1 \ll 1/2$ to control the interference to other users \cite{ratzer_crosstalk}. In
such scenarios, binary linear codes can not be applied directly since they can only induce the uniform
distribution. We refer to such channels as `channels with non-uniform inputs (NUI)'. The coding problem for such
channels remains open \cite{mceliece_talk}.

This problem was previously studied by Ratzer and Mackay \cite{ratzer_crosstalk} \cite{ratzer_sparse_code}. In
\cite{ratzer_sparse_code}, they focused on designing inverse Huffmann code type mappers to induce the desired
distribution. However, soft output decoding of the Huffman code is usually computationally complex  and,
further, the variable length nature of the mapping may incur catastrophic decoding errors. Alternatively, in
\cite{ratzer_crosstalk}, LDPC codes over GF(q) with deterministic mappers were used to induce the desired
non-uniform distribution. The main drawback of this scheme is that the decoding complexity for the nonbinary
LDPC code is significantly larger and the code optimization is very complicated.

In this paper, we first show that MLC using a set of binary linear codes and a deterministic mapper suffices to
achieve the information rate of the channel with NUI. The mapper, however, does not necessarily have to be
one-to-one. This scheme, discussed in Section \ref{sec:coding_nuic}, is shown to be optimal when the channel law
is known at the transmitter. Although an MLC scheme with binary inputs can only induce dyadic input
distributions, it is shown that via proper time sharing, the proposed MLC with a small number of layers can get
close to the channel information rate for an arbitrary $P_X$. Compared with the previous works, the proposed MLC
scheme not only has low complexity, but is theoretically justifiable as well.

As an important application of coding for channels with NUI, we consider the problem of rateless transmission
over the binary symmetric channel (BSC). In \cite{erez_rateless}, a simple layering, dithering (or interleaving)
and repeating based rateless scheme was proposed for AWGN channel. In this paper, we extend (non-trivially)
their results to the BSC case. Thanks to the degraded nature of the BSC, a similar layering scheme can be
applied without a rate loss. However, in order to perform layering, the number of 1s of the coded bits in each
layer must be constrained. This is precisely where the proposed MLC scheme can be applied. We further show that
repeating does not incur a rate loss in the low rate region over the BSC even for non-uniform inputs. Therefore,
rateless transmission over the BSC becomes possible by simply layering, interleaving and repeating the proposed
MLC block.

\section{Coding for Channels with Non-Uniform Inputs}
\label{sec:coding_nuic}

The problem of coding for channels with NUI can be dated back to Gallager. In \cite{gallager_book}, Gallager
showed that binary linear codes can achieve the capacity of any DMC:

\begin{thm}\label{thm:linear code for arbitrary DMS}
Binary linear codes can be used to achieve the capacity of an arbitrary discrete memoryless channel.
\end{thm}

We refer interested readers to \cite{gallager_book} for the detailed proof. The main result of this theorem says
that for any DMC, capacity can be achieved by a set of linear codes with a deterministic mapper under maximum
likelihood decoding (MLD). However, as suggested by Gallager, finding practical decoding algorithms is a
nontrivial problem. Note that in Gallager's proof, the key component, a deterministic mapper is used to induce
the desired channel input distribution to achieve the capacity of the DMC. The deterministic mapper can be
defined as follows:

\begin{defn}\label{dfn:dms}
A deterministic mapping is a function $f: W \rightarrow X$, where $W \in \{0, 1\}^m$ and $X \in {\cal X}$
(${\cal X}$ is the set of all possible channel input symbols).
\end{defn}

\begin{figure}
\begin{center}
\includegraphics[width=1.5in]{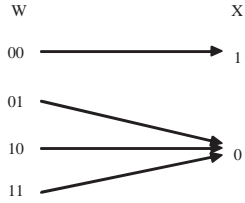}
\caption{An example of a deterministic mapper} \label{fig:deterministic}
\end{center}
\end{figure}

For example, consider the channel with NUI, i.e., $p_1 = P(X = 1) = 1/4$ and $p_0 = P(X = 0) = 3/4$. A possible
deterministic mapper is shown in Figure \ref{fig:deterministic}, where $W \in \{0, 1\}^{2}$ and $X \in \{0,
1\}$. Since $W$ is uniformly distributed, the mapper can induce the desired distribution on ${\cal X}$. Note
that using linear codes with a deterministic mapper, we can only obtain probabilities of the form $k/2^m$.
However, by increasing $m$, we can approximate the desired distribution arbitrarily well.

\begin{figure}
\begin{center}
\includegraphics[width=3.5in]{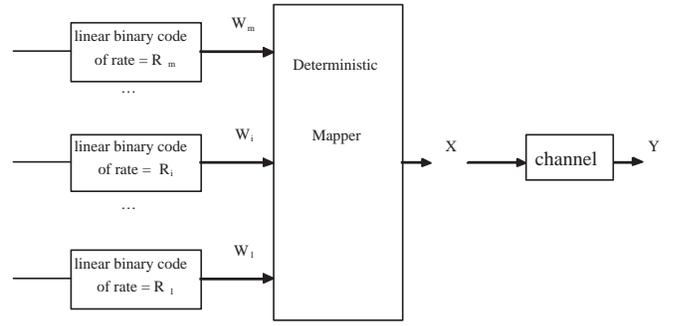}
\caption{System Model for the Proposed Coding Scheme for Constrained Input Channels} \label{fig:coding_nuic}
\end{center}
\end{figure}

\underline{Proposed~Scheme}: We first propose an MLC scheme to achieve the information rate of channels with
NUI. The diagram of the proposed MLC scheme is shown in Figure \ref{fig:coding_nuic} and the details of the
scheme is as follows: \textbf{Encoding}: In each layer, $W_i$ is encoded using a capacity achieving binary
linear code. The code rate of the $i^{th}$ layer is selected to be $R_i = I(W_i;Y|W_1, \cdots, W_{i-1})$. Then
we induce the desired distribution on ${\cal X}$ from $W = [W_1, \cdots, W_m]$ using a deterministic mapper as
suggested in Theorem \ref{thm:linear code for arbitrary DMS}. \textbf{Decoding}: At the decoder, we apply MSD:
$W_1$ is first decoded and then $W_2$ is decoded based on $Y$ and the decision of $W_1$ and so forth until $W_m$
is decoded based on $Y$ and all the decisions from $W_1$ to $W_{m-1}$.

Now, we show that the information rate of a channel with NUI can be achieved by the above MLC scheme.
\begin{thm}\label{thm:multilevel_capacity}
The proposed coding scheme can achieve the information rate of the DMC with NUI, i.e., $\sum_{i =
1}^{m}I(W_i;Y|W_1, \cdots, W_{i-1}) = I(X;Y)$.
\end{thm}
\begin{proof} \label{prf:asym_rateless_bsc}
We first show that the deterministic mapping from $W$ to  $X$ does not incur a rate loss. Note that $W
\rightarrow X \rightarrow Y$ forms a Markov chain. Expanding $I(W;Y, X)$ in two ways, we have:

\begin{align}
\label{align:channel_rule1} I(W;Y,X) &= I(W;X) + I(W;Y|X)\\
\label{align:channel_rule2}&= I(W;Y) + I(W;X|Y)
\end{align}

Due to the Markovian structure $I(W;Y|X) = 0$, thus, the mutual information between $W$ and $Y$ can be written
as:
\begin{align}
\label{align:determ1} I(W;Y) &= I(W;X) - I(W;X|Y)\\
\label{align:determ2} &= (H(X)-H(X|W)) - (H(X|Y)-H(X|W,Y))\\
\label{align:determ3} &= H(X) - H(X|Y) = I(X;Y)
\end{align}
(\ref{align:determ2}) to (\ref{align:determ3}) follows by the fact that $X$ is a function of $W$.

Hence, we can achieve the information rate using the proposed MLC. Since $W = [W_1, W_2, \cdots, W_m]$, the
mapping from ${W_1, W_2, \cdots, W_m}$ to $W$ is a bijection. According to the chain rule of mutual information,
we have:
\begin{equation}
\label{eqn:capacity} I(X;Y) = I(W;Y) = \sum_{i = 1}^{m}I(W_i;Y|W_1, \cdots, W_{i-1})
\end{equation}
\end{proof}

The proof generalizes the MLC proof in \cite{wachsmann_mlc}, where the mapping from $W$ to $X$ is a bijection.
However, here, $W$ does not need to be a deterministic function of $X$ , which suggests that the one-to-one type
mappers (e.g., inverse Huffman code \cite{ratzer_sparse_code}) are not required in order to achieve the
information rate. Essentially, what is needed is a deterministic mapper, which shapes the uniform distribution
obtained from the coded bits of the linear codes to be the desired channel NUI distribution. Besides, the
theorem implies that in each layer, standard binary linear codes, such as binary LDPC codes suffice.

\begin{figure}
\begin{center}
\includegraphics[width=3.0in]{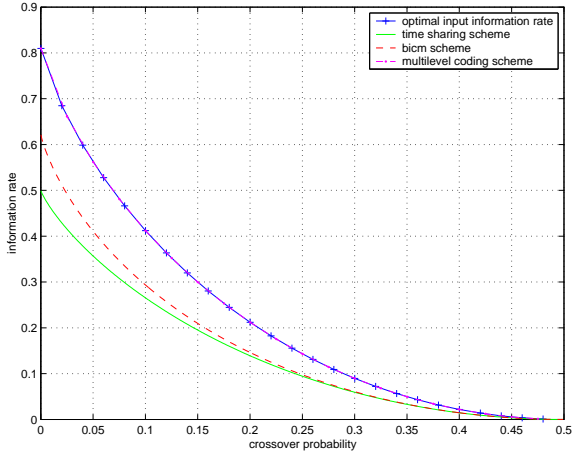}
\caption{Achievable Rate of Different Schemes over a BSC with $p_1$ = 1/4} \label{fig:bsc_rate}
\end{center}
\end{figure}

\underline{Example~1}: We give an example of the proposed MLC scheme over a BSC with NUI. In Figure
\ref{fig:bsc_rate}, the information rate is plotted as a function of the channel crossover probability $h$. The
probability of 1s at the channel input is fixed to be $p_1 = 1/4$. We can see that the proposed MLC can achieve
the information rate supported by the channel. In contrast, time sharing a linear code with 0s will incur a
significant rate loss. Besides, note that the mapper will introduce memory across the layers. Therefore,
bit-interleaved-coded-modulation (BICM) without iterative demodulation (the demapper generates bit-level soft
information for each layer by ignoring the correlation across the layers) also incurs a significant rate loss.
For Z channels, similar phenomenon is observed. Due to the page limit, the results are not shown here.

The input probabilities that can be induced are of the form $k/2^m$, i.e., dyadic fractional numbers. However,
any desired input probability $p_1$ can be approached by properly time sharing between two MLC schemes. For
instance, if we need $p_1 = 2/5$ at the channel input, we can time share between two MLC schemes with 2 layers,
one with $p_1 = 1/4$ and the other with $p_1 = 1/2$. The rate loss of the proposed MLC time sharing scheme is
usually small.

\begin{figure}
\begin{center}
\includegraphics[width=3.0in]{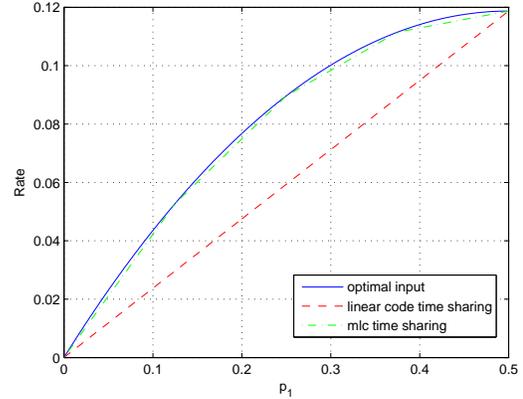}
\caption{Information Rate of a 3-Layer MLC Scheme over a BSC with h = 0.3} \label{fig:f03_compare}
\end{center}
\end{figure}

\underline{Example~2}: Consider the proposed MLC over for a BSC with NUI in Figure \ref{fig:f03_compare}. In
this case, the channel crossover probability is fixed $h = 0.3$ and the plot shows the change of the information
rate as a function of $p_1$. We can see that the simple scheme that time shares a linear code with 0s will incur
a substantial loss. To get close to the information rate, we may time share between two of the MLC schemes with
$m = 3$. As shown in Figure \ref{fig:f03_compare}, for any $k/8 \le p_1 \le (k+1)/8$, time sharing between the
MLC scheme with $p_1 = k/8$ and $p_1 = (k+1)/8$ achieves most of the information rate.

\section{Rateless Transmission Scheme over Binary Symmetric Channel using Layering and Repeating}
\label{sec:repeat optimality}

In this section, we present an application of the proposed scheme to rateless transmission over the BSC, which
is based on layering, interleaving and repeating. We show that in the low rate region, repeating preserves
information rate and therefore it can be used as a simple rateless scheme, which extends the result of
\cite{erez_rateless} to the BSC case. We then show that layering information does not incur a rate loss for the
BSC due to its degraded nature. As a result, in order to form layering, the coded bits in each layer have to be
non-uniformly distributed, where the proposed MLC scheme in Section \ref{sec:coding_nuic} becomes useful.

The problem of rateless transmission over BSC can be formulated as follows: suppose we want to communicate over
a BSC with an unknown but lower bounded crossover probability $h \ge h_{min}$. The capacity of this channel is
bounded by $0 \le C \le 1-H(h_{min})$. Since $h$ is unknown, the transmitter will first send a mother code of
rate $R_{max} = 1-H(h_{min})$ and then keep sending extra redundancies until the receiver gets enough
information to decode. The mother code together with any of the redundant suffix should be a good code such that
as long as right enough redundancies are collected, the receiver will be able to decode. Therefore, the proposed
scheme is called rateless, since it can work for a wide range of rates ($0 \le R \le 1-H(h_{min})$). For details
on rateless codes over the binary erasure channel (BEC), we refer interested readers to
\cite{Luby01a}\cite{shokrollahi_raptor}.

\begin{figure}
\begin{center}
\includegraphics[width=3.5in]{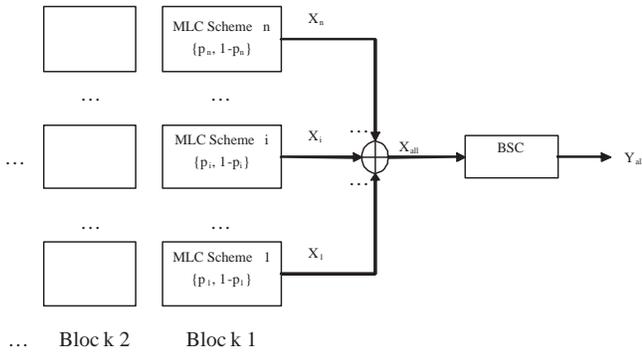}
\caption{the Proposed Rateless Scheme over the BSC} \label{fig:rateless}
\end{center}
\end{figure}

Here, we propose a rateless transmission scheme over the BSC based on layering, interleaving and repeating. The
structure of the rateless scheme over the BSC is shown in Figure \ref{fig:rateless}.

\underline{Proposed Scheme}: \textbf{Encoding}: In Block 1, each of the $i^{th}$ layer encodes its message into
coded bits obeying $\{p_i, 1-p_i\}$ Bernoulli distribution. The initial code rate of the $i^{th}$ layer $R_i$
should be selected such that $R_{max} = \sum_{i = 1}^{n}{R_i} = 1-H(h_{min})$. Then, the coded bits from each
layer are interleaved, since we have the NUI distribution, interleaving rather than dithering has to be used to
make sure that interference from other layers does not combine coherently when we combine the repeated blocks
(See \cite{erez_rateless} for details). All the interleaved layers are then stacked, i.e., bit wise XORed
together $X_{all} = \sum_{i = 1}^{n}{\oplus}{X_i}$ and transmitted through the BSC. If the receiver is not able
to decode, Block 2 is sent. That is all the coded bits of each layer are repeated, interleaved using a different
set of interleavers, then stacked together and transmitted through the channel again. The above procedure
continues until the receiver has got enough repeated blocks to decode, i.e., $mI(X_{all};Y_{all}) \ge R_{max}$.
\textbf{Decoding}: At the decoder, we first wait until enough number of blocks are collected. Then we apply MSD.
In the $n^{th}$ layer, we first generate the soft information of each coded bit from the repeated channel
outputs and use them to decode the $n^{th}$ layer's codeword. Then the decoded bits are subtracted and the
$(n-1)^{th}$ layer sees a clearer channel. We repeat the above decoding procedure until the $1^{st}$ layer is
decoded. Eventually, the information rate after $m$-time repetition is $R = R_{max}/m$.

%\begin{figure}
%\centerline{\epsfxsize=1.8in \epsfbox{rateless.eps} \epsfxsize=1.8in \epsfbox{degraded_bsc.eps}} \vspace*{-.1in}
%\caption{(a) the Proposed Rateless Scheme over the BSC, (b) Degraded Nature of the BSC} \label{fig:rateless}
%\end{figure}

As a prerequisite to show the optimality of the proposed scheme, we first give the following lemma:
\begin{lem}\label{lem:logsum}
Suppose a and b are constants, the function $f(x) = \log(b+ax)$ satisfies the following inequalities as $x
\rightarrow 0$
\begin{equation}
\label{eqn:f(x)} \log b + \frac{a}{b}x - \frac{1}{2}(\frac{a}{b}x)^2 \le \log(b+ax) \le \log b + \frac{a}{b}x
\end{equation}
\end{lem}
\begin{proof} \label{prf:logsum}
This lemma is immediate by considering the Taylor series expansion of $f(x)$ near $x = 0$.
\end{proof}

It is shown in \cite{shulman_thesis} that for channels with uniform input distribution, repeating preserves
information rate, in the low rate region. Here we extend this result to the BSC with NUI.

\begin{thm}\label{thm:low_rate_rateless}
Let $X$ be the channel input satisfying Bernoulli distribution $\{p, 1-p\}$. $X$ is transmitted through a BSC
with crossover probability $h$. Let $Y$ be the channel output and $Y^m$ be the m-time repetition of $X$ through
the BSC. When $p \rightarrow 0$, the information rate is preserved by repeating $X$ m times, i.e., $\lim_{p
\rightarrow 0}I(X;Y^m) = \lim_{p \rightarrow 0}mI(X;Y)$.
\end{thm}
\begin{proof} \label{prf:low_rate_rateless}
The channel output $Y$ will obey $\{p_y, 1-p_y\}$ Bernoulli distribution, where $p_y = p \otimes h =
p(1-h)+(1-p)h$.

We have the information rate:
\begin{align}
\label{align:bsc_rate1} I(X;Y) =& H(p \otimes h)-H(h)
%\\\label{align:bsc_rate2} =& -\left[(1-2h)p+h\right] \log{\left[(1-2h)p + h\right]} - \left[(1-h)-(1-2h)p\right]
%log{\left[(1-h)-(1-2h)p\right]} - H(h)
\end{align}

As p goes to zero, we have:
\begin{align}
\label{align:rateless} \lim_{p \rightarrow 0}{mI(X;Y)} =& \lim_{p \rightarrow
0}{mp(1-2h)\log{\frac{1-h}{h}}+o(p)}
\end{align}

On the other hand, we can derive the information combining of a repetition code over the BSC with NUI. Note that
$Y^m$ can be viewed as a vector channel output, there are $(m+1)$ types of channel outputs. Different outputs of
the same type are just different permutations and are statistically equivalent. We have the probability of the
$i^{th}$ type as:
\begin{align}
\label{align:multi-type1} p_i &= p h^i (1-h)^{m-i} + (1-p) h^{m-i}(1-h)^i\\
\label{align:multi-type2} &= p \left[h^i (1-h)^{m-i} - h^{m-i}(1-h)^i\right] + h^{m-i}(1-h)^i
\end{align}

Let $A_i = h^i (1-h)^{m-i} - h^{m-i}(1-h)^i$ and $B_i = h^{m-i}(1-h)^i$. Thus we have $p_i = A_i p + B_i$. Since
all the probabilities of all the channel outputs will sum up to be ``1''. We have:

\begin{equation}
\label{eqn:relationship1} \sum_{i = 0}^{m}{{{m}\choose{i}}(A_i p + B_i)} = 1
\end{equation}
Besides, since
\begin{equation}
\label{eqn:binomial_sum} \sum_{i = 0}^{m}{{{m}\choose{i}}B_i} = (h+1-h)^m = 1
\end{equation}
we have:
\begin{equation}
\label{eqn:relationship2} \sum_{i = 0}^{m}{{{m}\choose{i}}A_i} = 0
\end{equation}

The mutual information in the low rate region can therefore be written as:
\begin{align}
%\label{align:rep_combin1} I(X;Y^m) &= H(Y^m) - H(Y^m|X)\\
%\label{align:rep_combin2} H(Y^m) &= \sum_{i = 0}^{m} - {{m}\choose{i}}p_i\log{p_i}\\
%\label{align:rep_combin2} H(Y^m|X) &= mH(h)\\
\label{align:rep_combin3} \lim_{p \rightarrow 0}{I(X;Y^m)} &= \lim_{p \rightarrow 0}{\sum_{i = 0}^{m} - {{m}\choose{i}}A_i(\log{B_i}+\log{e})p + o(p)}\\
%\label{align:rep_combin4} &= \lim_{p \rightarrow 0}{\sum_{i = 0}^{m} - {{m}\choose{i}}A_i\log{B_i}p+o(p)}\\
%\label{align:rep_combin5} &= \lim_{p \rightarrow 0}{\sum_{i = 0}^{m} - {{m}\choose{i}}A_i\left[(m-i)\log{h}+i\log{(1-h)}\right]p+o(p)}\\
%\label{align:rep_combin6} &= \lim_{p \rightarrow 0}{\sum_{i = 0}^{m} - {{m}\choose{i}}A_i\left[m\log{h}+i\log{\frac{1-h}{h}}\right]p+o(p)}\\
\label{align:rep_combin7} &= \lim_{p \rightarrow 0}{p\log{\frac{1-h}{h}}\left[- \sum_{i = 0}^{m}{{m}\choose{i}}A_i i\right]+o(p)}\\
%\label{align:rep_combin8} &= \lim_{p \rightarrow 0}{p\log{\frac{1-h}{h}}\left[- \sum_{i = 0}^{m}{{m}\choose{i}} i h^i (1-h)^{m-i} + \sum_{i = 0}^{m}{{m}\choose{i}} i h^{m-i} (1-h)^i\right]+o(p)}\\
%\label{align:rep_combin9} &= \lim_{p \rightarrow 0}{p\log{\frac{1-h}{h}}\left[(1-h)m-fm\right]+o(p)}\\
\label{align:rep_combin10} &= \lim_{p \rightarrow 0}{pm(1-2h)\log{\frac{1-h}{h}}+o(p)}\\
\label{align:rep_combin11} &= \lim_{p \rightarrow 0}{mI(X;Y)+o(p)};
\end{align}
where (\ref{align:rep_combin3}) to (\ref{align:rep_combin7}) follows from Lemma \ref{lem:logsum} and some
straightforward manipulations. From (\ref{align:rep_combin11}), we can see that for very low rate, repeating
preserves information rate.
\end{proof}

\begin{figure}
\begin{center}
\includegraphics[width=3.5in]{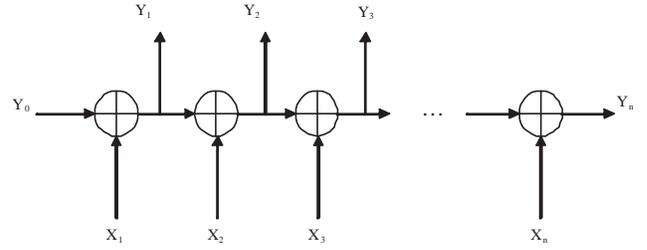}
\caption{Degraded Nature of BSC} \label{fig:degraded_bsc}
\end{center}
\end{figure}

In the proposed scheme, we use layering to drive the coding rate of each layer to the low rate region. We can
show that layering is lossless as follows:

\begin{thm}\label{thm:degraded_bsc}
For the BSC, layering does not incur any loss in information rate and MSD can be used to achieve the information
rate.
\end{thm}
\begin{proof} \label{prf:degraded_bsc}
The channel model of information layering over the BSC is shown in Figure \ref{fig:rateless}. Let the overall
stacked information as $X_{all} = \sum_{i = 1}^{n}{\oplus}{X_i}$ The overall channel output is $Y_{all} = Y_n$.
We have the following relationship:
\begin{align}
&I(X_1, X_2, \cdots, X_n;Y_n)\nonumber\\
\label{align:bsc_chain_rule1} &= \sum_{i = 1}^{n}{I(X_i;Y_n|X_{i+1}, \cdots, X_n)}\\
\label{align:bsc_chain_rule2} &= \sum_{i = 1}^{n}{I(X_i;Y_i)} = \sum_{i = 1}^{n}{(H(Y_i)-H(Y_{i-1}))}\\
\label{align:bsc_chain_rule3} &=  H(Y_n) - H(Y_{0})  = I(X_{all};Y_n)
\end{align}
From the above equations, (\ref{align:bsc_chain_rule3}) suggests that stacking does not incur a rate loss and
(\ref{align:bsc_chain_rule1}) suggests that the overall information rate can be achieved by MSD.
\end{proof}

It remains to show that the rate loss due to repeating does not accumulate as the number of layers increases.
Thus, as the number of layers goes to be large, the overall rate loss due to repeating is negligible.

\begin{thm}\label{thm:layering_rateless}
Using layering and repeating for rateless transmission is information lossless as long as the rate of each layer
is sufficiently small.
\end{thm}
\begin{proof} \label{prf:layering_rateless}
Let the total number of layers be $N$. For the $j^{th}$ layer, we have the channel input and output as $X_j$,
$Y_j$. The input probability $p_j$ and crossover probability $h_j$. Let the probability of 1 of the overall
information be $p_N$. For simplicity, let each layer has $p_i = p$. By recursion we have the following
relationship:
\begin{equation} \label{eqn:prob_relation}
p_j = p = \frac{1-(1-2p_N)^\frac{1}{N}}{2}
\end{equation}
(Note that $p_N = 1/2-\epsilon$, where $0 < \epsilon < 1/2$, since $p_i < 1/2$. As $N \rightarrow \infty$, $p_N$
can be made arbitrarily close to 1/2.)

The information rate per m-time channel use is $mI(X_j;Y_j)$, while the information rate of the m-time repeating
is $I(X_j;Y_j^m)$. We have the overall rate loss as:
\begin{equation}
\label{eqn:delta} \Delta = \lim_{N \rightarrow \infty}\sum_{j = 1}^{N}{\Delta_j} = \lim_{N \rightarrow \infty}
\sum_{j = 1}^{N}{\left[mI(X_j;Y_j) - I(X_j;Y_j^m)\right]} = 0
\end{equation}

Note that the information rate of each layer can be made arbitrarily small such that Lemma \ref{lem:logsum}
holds. Thus the information rate per m-time channel use is upper bounded by:
\begin{equation} \label{eqn:upper_bound}
\lim_{p \rightarrow 0}{mI(X_j;Y_j)} \le \lim_{p \rightarrow 0}{mp(1-2h_j)\log{\frac{1-h_j}{h_j}}}
\end{equation}
On the other hand, following Lemma \ref{lem:logsum} we have the following inequalities:
\begin{align}
\label{align:logsum1} \log(B_i+A_i p) &\ge \log{B_i}+\frac{A_i}{B_i}p-\frac{1}{2}(\frac{A_i}{B_i})^2p^2\\
\label{align:logsum2} \log(1-B_i-A_i p) &\ge
\log{(1-B_i)}-\frac{A_i}{1-B_i}p-\frac{1}{2}(\frac{A_i}{1-B_i})^2p^2
\end{align}
Note that $A_i$ and $B_i$ do not depend on the number of layers $N$ and consequently they are bounded. Thus,
plugging (\ref{align:logsum1}) (\ref{align:logsum2}) into (\ref{align:rep_combin3}), we have a lower bound of
the information rate of the m-time repetition $I(X_i;Y_i^m)$ as:
\begin{equation} \label{eqn:lower_bound}
\lim_{p \rightarrow 0}{I(X_j;Y_j^m)} \ge \lim_{p \rightarrow 0}{mp(1-2h_j)\log{\frac{1-h_j}{h_j}} - c m p^2}
\end{equation}
where $c$ is constant.

Combining (\ref{eqn:upper_bound}) and (\ref{eqn:lower_bound}) the rate loss can be bounded by:
\begin{equation} \label{eqn:gap_upper_bound}
\Delta_j = mI(X_j;Y_j) - I(X_j;Y_j^m) \le c m p^2
\end{equation}

Consequently, we have the overall information loss as:
\begin{align}
\label{align:delta_zero1} \Delta &= \lim_{N \rightarrow \infty}\sum_{j = 1}^{N}{\Delta_j}\\
\label{align:delta_zero2} &\le c m \lim_{N \rightarrow \infty}{\left[\frac{1-(1-2p_N)^\frac{1}{N}}{2}\right]^2N}
\propto \lim_{N \rightarrow \infty} \frac{1}{N} = 0
\end{align}
\end{proof}

Note that $p_i < 1/2$ for all the layers, i.e., each layer will have to have NUI distribution, since otherwise
the interference seen by the upper layers will have crossover probability 1/2. Thus, in order to perform the
rateless transmission, we need codes for channels with NUI, where the proposed MLC scheme discussed in the
previous section becomes useful.

%\section{Conclusion}
%\label{sec:conclusion}
%We considered coding schemes for channels with NUI and its application to rateless transmission over the BSC.

\bibliographystyle{unsrt}
\bibliography{./jiang}
\end{document}